\documentclass[12pt]{article}

\usepackage{amssymb}
\usepackage{amsmath}
\usepackage{amscd}
\usepackage{latexsym}

\newcommand{\be}{\begin{equation}}
\newcommand{\ee}{\end{equation}}

\newcommand{\dlt}{\delta}

\newcommand{\bt}{\beta}

\newcommand{\ep}{\varepsilon}
\newcommand{\al}{\alpha}

\newcommand{\cM}{{\cal M}}
\newcommand{\cA}{{\cal A}}
\newcommand{\cL}{{\cal L}}
\newcommand{\cD}{{\cal D}}

\begin{document}

\begin{center}

{\Large{\bf Entanglement production in quantum decision making} \\ [5mm]

V.I. Yukalov$^{1,2}$ and D. Sornette$^{1,3}$} \\ [3mm]

{\it $^1$ETH Z\"urich, Z\"urich, Switzerland \\ [3mm]

$^2$Joint Institute for Nuclear Research, Dubna, Russia \\ [3mm]

$^3$Swiss Finance Institute, c/o University of Geneva, Geneva, Switzerland}

\end{center}

\vskip 2cm

\begin{abstract}

The quantum decision theory (QDT) introduced recently
is formulated as a quantum theory of measurement. It describes
prospect states represented by complex vectors
of a Hilbert space over a prospect lattice. The prospect operators,
acting in this space, form an involutive bijective algebra. A
measure is defined for quantifying the entanglement produced by the
action of prospect operators. This measure characterizes the level
of complexity of prospects involved in decision making. An explicit
expression is found for the maximal entanglement produced by the
operators of multimode prospects.
\end{abstract}

\vskip 2cm
{\bf pacs}: 03.67.Hk, 89.75.-k, 89.70.Hj

\newpage

\section{Introduction}\label{int1}

Entanglement is a quantum property that is very important for
quantum information processing and quantum computing [1,2]. It
is one of the key features for creating artificial intelligence
based on quantum rules. Our recent formulation of Quantum
Decision Theory [3,4] is based on the recognition that entanglement
is also a characteristic property of human decision making.
Indeed, any real decision making procedure deals with
composite prospects, composed of many intended actions, which
produce naturally entanglement due to correlations 
between particular actions. These correlations need only to be
subjectively felt in the brain of the decision maker to affect
his/her choices, strongly colored by the emotional
effects and aversion to risk and uncertainty.
This has led us to use the mathematics of
quantum theory, in order to develop a decision theory
of non-quantum objects, such as human decision makers [3,4]. 
This approach can be also applied 
to physical devices of quantum information processing [5]. 

With the understanding that almost any decision procedure involves
entanglement, it then becomes necessary to quantify in some way the level
of produced entanglement. While we have modeled the phenomenon
of entanglement to explain and quantify many classical paradoxes arising
in standard utility theory, a systematic measurement of entanglement
has not been developed
in the previous publications on QDT [3-5]. It is the aim of
the present paper to analyze the problem of entanglement
production that can be generated in the process of decision
making.

\section{Algebra of prospect operators}\label{sec2}

In order to construct the mathematics of QDT, we employ the techniques
of quantum theory of measurement [6,7], with the specifications
appropriate for describing the process of decision making.
The primary objects of a decision procedure are the intended
actions, whose totality forms the {\it action ring}
\be
\label{eq1}
\cA = \{ A_n: \; n =1,2,\ldots, N\} \; .
\ee
Each action, generally, is composed of several representations,
called {\it action modes},
\be
\label{eq2}
A_n = \bigcup_{\mu=1}^{M_n} A_{n\mu} \qquad
( A_{n\mu} A_{n\nu} =\dlt_{\mu\nu}) \; .
\ee
A prospect is a conjunction of several actions,
\be
\label{eq3}
\pi_j = \bigcap_{n=1}^N A_{j_n} \qquad (A_{j_n}\in\cA) \; .
\ee
It can be simple, if each action is represented by a single mode,
or composite, when there is at least one composite action in the
conjunction. The family of all prospects forms a lattice
\be
\label{eq4}
\cL = \{ \pi_j: \; j = 1,2, \ldots, N_L \} \; ,
\ee
endowed with the binary operations $<$ (``less preferred than''), 
$>$ (``more preferred than''), and $=$ (``equivalent to'' or ``indifferent 
with''), so that each two prospects from ${\cal L}$ are connected either 
as $\pi_i\leq\pi_j$ or as $\pi_i \geq \pi_j$, or as $\pi_i = \pi_j$. 
Elementary prospects are defined as simple disjoint prospects
\be
\label{eq5}
e_\al =  \bigcap_{n=1}^N A_{i_n \mu_n}  \qquad
(e_\al e_\bt = \dlt_{\al\bt} ) \; ,
\ee
containing only single modes and labelled with a multi-index
$\alpha = \{i_n, \mu_n : n = 1,2,...,N_L\}$. The cardinality of the
set $\{\alpha\}$ is ${\rm card}\;\{\alpha\} = \prod_{n=1}^N M_n$.

Each mode $A_{n \mu}$ corresponds to a mode state $|A_{n \mu}>$,
which is a complex-valued function with an orthonormalized scalar
product $<A_{n\mu}|A_{n\nu}> = \delta_{\mu\nu}$. The closed
linear envelope, spanning all mode states, is the {\it mode space}
\be
\label{eq6}
\cM_n =  {\rm Span} \{ | A_{n\mu} >\; : \;
\mu =1,2,\ldots,M_n  \} \; ,
\ee
with the dimensionality ${\rm dim}\;\cM_n = M_n$.

An elementary prospect $e_\alpha$ corresponds to a
{\it basic state} $|e_\alpha>$, which is a complex function
\be
\label{eq7}
|e_\al > \; = \; | A_{i_1\mu_1} A_{i_2\mu_2} \ldots
A_{i_N\mu_N} > \; = \; \bigotimes_{n=1}^N |A_{i_n\mu_n} >  \; .
\ee
The basic states are orthonormalized, such that
$<e_\alpha|e_\beta> = \delta_{\alpha \beta}$. The closed linear
envelope, spanning all basic states, is the {\it mind space}
\be
\label{eq8}
\cM = {\rm Span} \{ |e_\al>\; : \; \al \in \{ \al\} \; \} =
\bigotimes_{n=1}^N \cM_n \; ,
\ee
whose dimensionality is ${\rm dim}\;{\cal M}={\rm card}\{\alpha\} =
\prod_{n=1}^N M_n$.

A prospect $\pi_j$ is represented by a prospect state $|\pi_j>$,
which is a member of the mind space $\cM$. That is, it can be
expanded over the basic states,
\be
\label{eq9}
| \pi_j> \; = \; \sum_\al b_j(e_\al) | e_\al > \; , \qquad
b_j(e_\al) \; = \; < e_\al | \pi_j > \; .
\ee
The prospect operator is defined as
\be
\label{eq10}
\hat P (\pi_j) =  |\pi_j > < \pi_j |  \; ,
\ee
with the condition that the sum
$$
\sum_{j=1}^{N_L} \hat P(\pi_j) = \hat 1_\cM 
$$
over the prospect lattice is a unity operator in the weak sense, 
with respect to the matrix element over a fixed strategic state 
$|s>$ characterizing the considered decision maker. That is, the 
above equality has to be understood as the equality on the average
$$
\sum_{j=1}^{N_L} \; < s | \hat P(\pi_j)| s > \; = \; 1 \; ,
$$
with $|s>$ being the given strategic state [3-5].
The involution is given by the Hermitian conjugation. By their
definition, the prospect operators (\ref{eq10}) are self-adjoint.
Hence, the family of the prospect operators forms an involutive
bijective algebra.

\section{Prospect produced entanglement}\label{sec3}

We now introduce a measure of the
amount of entanglement produced by a prospect operator.
To understand how the
prospect operators entangle the mind states, we need first to
classify the latter into entangled or disentangled states. A
mind state is disentangled if it can be represented as a product
state or {\it factor state}
\be
\label{eq11}
| f > \; \equiv \; \bigotimes_{n=1}^N \; |f_n> \qquad
( |f_n>\; \in \cM_n ) \; .
\ee
The ensemble of all such factor states forms the
{\it disentangled set}
\be
\label{eq12}
\cD \equiv \left \{ | f > \; = \;
\bigotimes_{n=1}^N \; | f_n> \; \in \; \cM \right \} \; ,
\ee
hence, ${\cal D} \subset {\cal M}$. The complement
${\cal M}\setminus{\cal D}$ composes the entangled set.

The entangling properties of the prospect operator can be
understood by comparing its action with that of its nonentangling
counterpart composed as a product of the partially traced
prospect operators
\be
\label{eq13}
\hat P_n(\pi_j) \; \equiv \;
{\rm Tr}_{\{ \cM_m :\; m\neq n \} } \; \hat P(\pi_j) \; ,
\ee
where the trace is over all $\cM_m$ except $m=n$. The
{\it nonentangling prospect operator}
\be
\label{eq14}
\hat P^\otimes (\pi_j) \equiv
\frac{{\rm Tr}_\cM \hat P(\pi_j)}
{{\rm Tr}_\cM\bigotimes_{n=1}^N \hat P_n(\pi_j) } \;
\bigotimes_{n=1}^N \; \hat P_n(\pi_j)
\ee
is the product of the partially traced operators (\ref{eq13}), defined
so that to preserve the normalization condition
\be
\label{eq15}
{\rm Tr}_\cM \; \hat P(\pi_j) =
{\rm Tr}_\cM \; \hat P^\otimes (\pi_j) \; .
\ee
The following equalities hold for the traces:
$$
{\rm Tr}_\cM \; \hat P(\pi_j) =
{\rm Tr}_{\cM_n} \; \hat P_n (\pi_j) = \sum_\al | b_j(e_\al)|^2 \; ,
$$
\be
\label{eq16}
{\rm Tr}_\cM \bigotimes_{n=1}^N \; \hat P_n(\pi_j) =
\prod_{n=1}^N {\rm Tr}_{\cM_n} \; \hat P_n(\pi_j) =
\left ( \sum_\al |b_j(e_\al)|^2 \right )^N \; .
\ee
As a result, the nonentangling operator (\ref{eq14}) takes the form
\be
\label{eq17}
\hat P^\otimes(\pi_j) = \left ( \sum_\al \;
|b_j(e_\al)|^2 \right )^{1-N} \;
\bigotimes_{n=1}^N \; \hat P_n(\pi_j) \; .
\ee

The entangling properties of the prospect operator (\ref{eq10}) are
the most clearly pronounced in the action of the prospect operator
on the disentangled set (\ref{eq12}). On this set, we may define the
restricted norm
\be
\label{eq18}
|| \hat P(\pi_j) ||_\cD \equiv \sup_{|f>\; \in\; \cD} \;
\frac{|< f| \hat P(\pi_j)| f > | }{ | < f| f> | } \; ,
\ee
which is a kind of a subnorm [8,9]. In particular, we have
$$
|| \hat P(\pi_j) ||_\cD = \sup_\al | b_j(e_\al)|^2 \; ,
\qquad
|| \hat P^\otimes(\pi_j) ||_\cD =
\frac{ ||\bigotimes_{n=1}^N \hat P_n(\pi_j)||_\cD}
{\left ( \sum_\al | b_j(e_\al)|^2\right )^{N-1} } \; ,
$$
\be
\label{eq19}
||\bigotimes_{n=1}^N \;  \hat P_n(\pi_j) ||_\cD =
\prod_{n=1}^N \; || \hat P_n(\pi_j) ||_{\cM_n} \; .
\ee

The measure of entanglement production [10-12], generated by
the prospect operator (\ref{eq10}), is defined as
\be
\label{eq20}
\ep(\pi_j) \equiv \log \; \frac{ || \hat P(\pi_j) ||_\cD}
{|| \hat P^\otimes(\pi_j) ||_\cD } \; ,
\ee
where the logarithm can be defined with respect to any base, say, to the base two.
Taking into account Eqs. (\ref{eq16}) and (\ref{eq19}) yields
\be
\label{eq21}
\ep(\pi_j) = \log \;
\frac{\sup_\al |b_j(e_\al)|^2\left(\sum_\al | b_j(e_\al)|^2\right)^{N-1}}
{\prod_{n=1}^N \; || \hat P_n(\pi_j) ||_{\cM_n} } \; .
\ee

In order to evaluate the maximal entanglement that could be
generated by the prospect operators, we should consider the
maximally entangled prospect states $|\pi_j>$. For the simplest
case of two actions with two modes each, the maximally
entangled state is of the Bell type
$$
| \pi_B > \; = \; b_1 | A_{11} A_{21} > +
b_2| A_{12} A_{22} > \; .
$$
When there are $N$ two-mode actions, the prospect state is
a multicat state
$$
| \pi_C > \; = \; b_1 | A_{11} A_{21} \ldots A_{N1} > +
b_2 | A_{12} A_{22} \ldots A_{N2} > \; .
$$
The general case of a maximally entangled state is a
multimode state
\be
\label{eq22}
| \pi_M > \; = \; \sum_{\mu=1}^M \;
b_\mu | A_{1\mu} A_{2\mu} \ldots A_{N\mu} > \; ,
\ee
corresponding to $N$ actions with $M$ modes. The related
prospect operator, characterizing the multimode prospect, is
\be
\label{eq23}
\hat P(\pi_M) = | \pi_M > < \pi_M | \; .
\ee
For this operator, we have
$$
|| \hat P(\pi_M) ||_\cD =
|| \hat P_n(\pi_M) ||_{\cM_n} = \sup_\mu |b_\mu|^2 \; .
$$
Therefore, measure (\ref{eq21}) transforms into
\be
\label{eq24}
\ep(\pi_M) = (N-1) \log \;
\frac{\sum_{\mu=1}^M | b_\mu|^2} {\sup_\mu | b_\mu|^2} \; .
\ee
Expression (\ref{eq24}) acquires its maximal value when all modes
are equally probable, such that $|b_\mu| = |b| = const$. Then the
measure of entanglement production (\ref{eq24}) becomes
\be
\label{eq25}
\ep(\pi_M) = (N-1) \log M \; .
\ee

In this way, the maximal measure of entanglement that can be
generated by a prospect, consisting of $N$ actions,
corresponds to the case when all actions possess the same
number of equiprobable modes $M$. Then the measure of
entanglement production is proportional to the number of
actions and logarithmically depends on the number of modes.

\section{Conclusion}

We have formulated quantum decision theory as the
theory of quantum measurements. We have suggested a method
for evaluating the entanglement generated by the prospect
operators in QDT. The measure of entanglement production
depends on the structure of the prospects involved. This
measure can be employed for quantifying the complexity of
prospects in decision theory. It can also be used for
characterizing the complexity of operations in quantum
information processing.

The most effective information processing requires a high level 
of produced entanglement [1,2,13]. We have shown that the maximal 
entanglement production is characterized by formula (\ref{eq25}).
The developed theory can be applied to nonquantum decision makers 
[3,4] as well as to quantum objects of different physical nature 
[5]. Except spin systems, multimode coherent states of trapped 
atoms [14,15] seem to be good candidates for realizing 
quantum-information processing devices.

\vskip 0.5cm
{\large{\bf Aknowledgements}}:

\vskip 2mm

The authors are grateful to the organizers of the
International Colloquium on Group Theoretical Methods in
Physics, Yerevan, Armenia, 2008, for providing an
opportunity to present the results of this work.
Financial support from ETH Zurich (Swiss Federal Institute of
Technology) and Russian Foundation for Basic Research is
appreciated.


\vskip 2cm

\begin{thebibliography}{99}
\bibitem{1}
C.P. Williams and S.H. Clearwater, {\it Explorations in Quantum
Computing} (Springer, New York, 1998).

\bibitem{2}
M.A. Nielsen and I.L. Chuang, {\it Quantum Computation and Quantum
Information} (Cambridge University, New York, 2000).

\bibitem{3}
V.I. Yukalov and D. Sornette, arXiv.org.0802.3597 (2008).

\bibitem{4}
V.I. Yukalov and D. Sornette, ssrn.com/abstract=1263853 (2008).

\bibitem{5}
V.I. Yukalov and D. Sornette, Phys. Lett. A {\bf 372}, 6867
(2008).

\bibitem{6}
J. Von Neumann, {\it Mathematical Foundations of Quantum Mechanics}
(Princeton University, Princeton, 1955).

\bibitem{7}
P.A. Benioff, J. Math. Phys. {\bf 13}, 908 (1972).

\bibitem{8}
 M. Goldberg and W.A. Luxemburg, Linear Algebra Appl. {\bf 307},
89 (2000)

\bibitem{9}
M. Goldberg, Electron. J. Linear Algebra {\bf 17}, 359 (2008).

\bibitem{10}
V.I. Yukalov, Phys. Rev. Lett. {\bf 90}, 167905 (2003).

\bibitem{11}
V.I. Yukalov, Mod. Phys. Lett. B {\bf 17}, 95 (2003).

\bibitem{12}
V.I. Yukalov, Phys. Rev. A {\bf 68}, 022109 (2003).

\bibitem{13}
A. Vedral, Rev. Mod. Phys. {\bf 74}, 197 (2002).

\bibitem{14}
V.I. Yukalov and E.P. Yukalova, Laser Phys. {\bf 16}, 354 (2006).

\bibitem{15}
V.I. Yukalov and E.P. Yukalova, Phys. Rev. A {\bf 73}, 022335 (2006).

\end{thebibliography}
\end{document}